\documentclass[aps, prb, twocolumn,floatfix, showpacs, 10pt]{revtex4}
\usepackage{amsmath}
\usepackage{amsfonts}
\usepackage{amsbsy}
\usepackage{amssymb}
\usepackage{graphicx}
\usepackage{textcomp}
\usepackage[caption=false,singlelinecheck=false]{subfig}
\usepackage{xcolor}
\usepackage{mathrsfs}
\usepackage{bm}
\usepackage[colorlinks,linkcolor=red,citecolor=blue,filecolor=magenta,
urlcolor=cyan]{hyperref}

\allowdisplaybreaks

\begin{document}
\title{Effects of Interaction in the Hofstadter regime of
the honeycomb lattice}

\author{Archana Mishra}

\author{S. R. Hassan}

\author{R. Shankar}

\affiliation{The Institute of Mathematical Sciences, C.I.T. Campus, Chennai 600 
113, India}

\date{\today}
%%%%%%%%%%%%%%%%%%%%%%%%%%%%%%%%%%%%%%%%%%%%%%%%%%%%%%%%%%%%%%%%%%%%%%%%%%%%%%%%%%%%%%%%%%%%%%%%%
%%%%%%%%%%%%%%%%%%%%%%%%%%%%%%%%%%%%%%%%%%%%%%%%%%%%%%%%%%%%%%%%%%%%%%%%%%%%%%%%%%%%%%%%%%%%%%%%%
\begin{abstract}
We investigate phases of spinless fermions on the honeycomb lattice with
nearest neighbor interaction in the Hofstadter regime. The interaction induces
incompressible nematic and ferri-electric phases with broken translation
symmetry. Some of the transitions are accompanied by changes in the Hall
conductivity. We study pair correlations and show that the quantum metric,
averaged over the Brillouin zone, characterizes the shape of the pair
correlation function.
\end{abstract}
%%%%%%%%%%%%%%%%%%%%%%%%%%%%%%%%%%%%%%%%%%%%%%%%%%%%%%%%%%%%%%%%%%%%%%%%%%%%%%%%%%%%%%%%%%%%%%%%%%%
%%%%%%%%%%%%%%%%%%%%%%%%%%%%%%%%%%%%%%%%%%%%%%%%%%%%%%%%%%%%%%%%%%%%%%%%%%%%%%%%%%%%%%%%%%%%%%%%%%%%
\maketitle
%%%%%%%%%%%%%%%%%%%%%%%%%%%%%%%%%%%%%%%%%%%%%%%%%%%%%%%%%%%%%%%%%%%%%%%%%%%%%%%%%%%%%%%%%%%%%%%%%%%%%
\section{Introduction}

Two dimensional electron systems in a periodic potential in the presence of a
magnetic field are characterized by two length scales, the periodicity of the
potential and the magnetic length. The regime where these two length scales are
comparable exhibits very interesting phenomena like the Hofstadter butterfly
\cite{hofstadter1976}. Recently this regime has been accessed experimentally
with observations of a stable Hofstadter spectrum in graphene superlattices
\cite{ponomarenko2013,dean2013,hunt2013, yu2014,yu12014} and realization of the
Hofstadter Hamiltonian in cold atoms systems
\cite{aidelsburger2013,miyake2013}. This has motivated us to investigate the
effects of repulsive interactions in this regime. The interactions are expected
to induce charge ordering in the ground state which may spontaneously break the
translational symmetry of the system. Consequently, one of the length scales,
namely the periodicity, can change. This could change the fractal structure.
Therefore, we investigate translational symmetry breaking in this system.
Interaction induced translational symmetry breaking phases have been studied in
the honeycomb lattice in the absence of magnetic field \cite{weeks2010,castro2011,castro2013}.
Effects of interactions on the Hofstadter butterfly have been discussed
previously \cite{gudmundsson1995,doh1998,czajka2006,chakraborty2013,apalkov2014}; these works
do not consider translational symmetry breaking.

We study a system of spinless fermions on the honeycomb lattice in the
Hofstadter regime with nearest neighbor repulsive interactions using mean field
theory. We restrict ourselves to the cases when the Fermi level lies in a gap
and thus expect the mean field approximation to be good. The ground state of
the non-interacting system has all the symmetries of the Hamiltonian and
non-zero Hall conductivity. Our studies reveal various interesting phases as
the strength of the interaction is increased. We get a first order Landau
transition to a phase with broken translational and rotational symmetries.
This phase is an incompressible nematic phase characterized by an electric
quadrupole moment of the ground state charge distribution. The transition is
accompanied by a change in the Hall conductivity.  In some cases, on further
increasing the interaction, there is another first order Landau transition to
an incompressible ferri-electric phase where the inversion symmetry is also
broken and the system develops an electric dipole moment. 

Nematic phases in homogeneous quantum Hall systems have previously been
theoretically studied \cite{koulakov1996,haldane2011,you2014,maciejko2013} and
experimentally observed \cite{lilly1999,xia2011,liu2013} in fractional quantum
Hall systems. Our results show that they occur in the Hofstadter regime also. 

The nematic order parameter has been related to the quantum metric
\cite{maciejko2013, haldane2011}.  This motivates us to investigate the
structure of the anisotropic phases by studying pair correlations from the
point of view of the quantum geometric approach to insulating states
\cite{resta2011}. We find that the shape and the extent of the pair
correlations can be exactly related to the quantum metric in the momentum space
\cite{resta2011}, averaged over the Brillouin zone (BZ) \cite{resta2006}. This result relates
the momentum space quantum metric of systems in a periodic potential to the
real space metric introduced by Haldane \cite{haldane2011} in homogeneous
quantum Hall systems.

The remaining part of the paper is organized as follows. We discuss the
interacting Hofstadter Hamiltonian and its symmetries in section \ref{sec:II}.
The noninteracting physics of this model and the band  topology is illustrated
here.  The mean field approximation used to solve the interacting Hamiltonian
is described in section \ref{sec:III}.  We give explicit description of the
complex phases and the  phase transitions obtained from solving the self
consistency equations in section \ref{sec:IV}. In section \ref{sec:V}, we
discuss the geometry of the ground state and give a relation between
pair correlation function and the 
quantum metric. We summarize and discuss the results in section \ref{sec:VI}.
The Appendices \ref{sec:A1} and \ref{sec:A2} contain the description of  Chern numbers
at half filling and the energy band diagrams in
the symmetric and nematic phases respectively.

\section{\label{sec:II}Model and Symmetries}
The model Hamiltonian we study is,
\begin{equation} 
\label{ham1} H=-t\sum_{\langle ij\rangle}
\left(c^\dagger_i e^{i\frac{e}{\hbar}A_{\langle ij\rangle}}c_j
+h.c\right) +V\sum_{\langle ij\rangle}n_in_j, 
\end{equation} 
%%%%%%%%%%%%%%%%%%%%%%%%%%%%%%%%%%%%%%%%%%%%%%%%%%%%%%%%%%%%%%%%%%%%%%%%%%%%%%%%%%%%%%%%%%%%%%%%%%%%%%%%%%%%%
where $c_{i}(c_{i}^{\dag})$ is the annihilation (creation) operator for
electrons at site $i$ on the honeycomb lattice, $n_{i}$ is the number density
operator, $t$ is the nearest  neighbor hopping parameter and $V$ is the nearest
neighbor interaction strength.
We consider $t=1$ and $V$ is in units of the
hopping matrix element. $A_{\langle ij\rangle}$ are the gauge fields on the
nearest neighbor links such that the magnetic flux passing through
each plaquette is $\phi=\frac{1}{q}\frac{h}{e}$ where $q$ is an integer. 

\begin{figure}[hbtp]
\begin{center}
\includegraphics[width=0.33\textwidth]{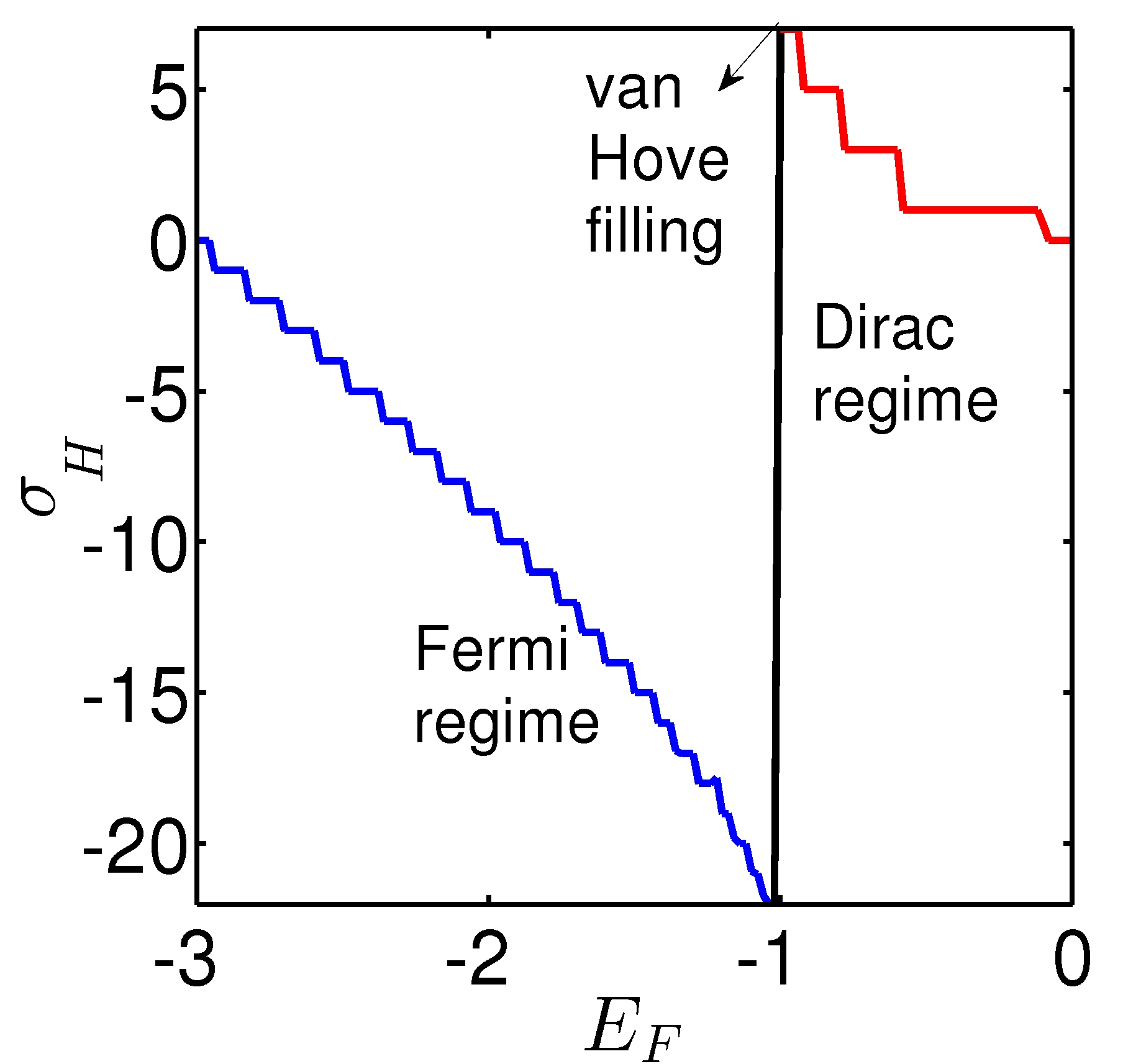}
\caption{\label{fig0}(Color online) Hall conductivity vs Fermi energy plot for $q=30$.
The Fermi regime, van Hove filling and Dirac regime are shown in this figure.}
\end{center}
\end{figure}

For these flux values, it has been shown \cite{hatsugai2006} that the
non-interacting theory has three regimes of electron densities. The dilute
limit is called the Fermi regime where each filled band has Hall conductivity,
$ \sigma_H=-e^2/h$.  Thus, $\sigma_H=-me^2/h$ when $m$ bands are filled as is
the case for non-relativistic systems in the continuum. At a certain filling,
$m^*$, there is a band with a very large Chern number where the Hall
conductivity changes sign when it is completely filled. This band lies in the
energy region of the van Hove singularity of the system in the absence of the
magnetic field. We refer to the $m^*$ filling as the van Hove filling. Pairs of
bands get degenerate (at large $q$) on further filling. The Hall conductivity
changes in steps of $2e^2/h$ when the Fermi level lies in the gap. This is
called the Dirac regime. Fig.~\ref{fig0} shows the plot for Hall
conductivity vs Fermi energy showing  the three different regimes for $q=30$.
The blue line is for the Fermi regime and the red line is in the  Dirac
regime.

In this paper we concentrate on $q=3$  which is the simplest case where, in the
non-interacting system, the bands are well separated with energy gaps, except
for the middle bands which touch at 6 Dirac points.  The Chern numbers of the
$m=1,2$ and $3$ bands are $(-1,2,1)$. Consequently, the Hall conductivities of
the system with one, two and three bands filled are $\sigma_H=-e^2/h,~e^2/h$
and 0 respectively. Going by the behavior of the large $q$ systems discussed
above we refer to these three fillings as the Fermi, van Hove and Dirac
fillings.

\begin{figure}[hbtp]
\begin{center}
\includegraphics[width=0.33\textwidth]{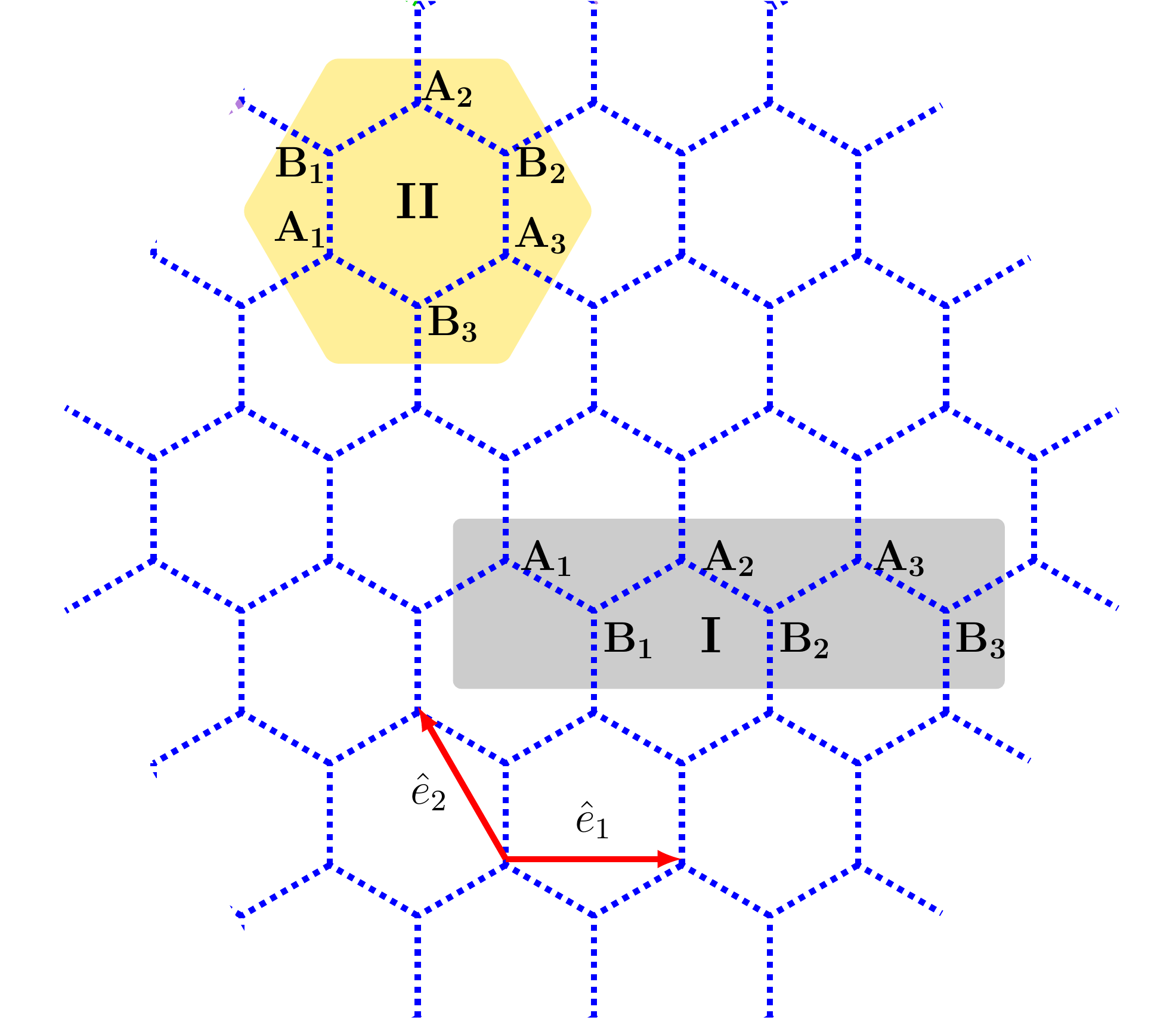}
\caption{\label{fig1}(Color online) Honeycomb lattice in a magnetic
field with flux $\phi=\frac{h}{3e}$ passing through each plaquette. 
$A$ and $B$ are the two sublattices. We consider two possible choices of unit cell:
linear choice ({\bf I}) shown as a gray rectangle and hexagonal choice ({\bf II})
shown as a yellow hexagon. $(A_1, B_1, A_2, 
B_2, A_3, B_3)$ are sublattices of these magnetic unit cells. ${\hat
e}_{1}$ and ${\hat e}_{2}$ represent the basis vectors of the original
lattice.}
\end{center}
\end{figure}

The Hamiltonian is invariant under magnetic translations, $\tau_1$ and $\tau_2$
which are along $\hat e_1$ and $\hat e_2$ directions respectively shown in
Fig.~\ref{fig1}. These magnetic translation operators do not commute with each
other, $\tau_1\tau_2\tau_1^{-1}\tau_2^{-1}=e^{i\frac{2\pi}{3}}$. Thus we need
to choose a magnetic unit cell consisting of three original unit cells to
implement the Bloch theory. There are two ways of doing this: the linear unit
cell denoted as unit cell choice {\bf{I}} with basis vectors $3\hat e_1,~\hat
e_2$ as shown by the rectangular region shaded in gray in Fig.~\ref{fig1} and
the hexagonal unit cell denoted as unit cell choice {\bf{II}} with basis
vectors $\hat e_2-\hat e_1,~\hat e_2+2\hat e_1$ as shown by the hexagonal
region shaded in yellow in Fig.~\ref{fig1}.  Unit  cell  choice {\bf II} is a
rotational symmetric choice where each sublattice is surrounded with three
different sublattices unlike unit cell  choice {\bf I}.

%%%%%%%%%%%%%%%%%%%%%%%%%%%%%%%%%%%%%%%%%%%%%%%%%%%%%%%%%%%%%%%%%%%%%%%%%%%%%%%%%%%%%%%%%%%%%%%%%%%%%%%%%%%%
%%%%%%%%%%%%%%%%%%%%%%%%%%%%%%%%%%%%%%%%%%%%%%%%%%%%%%%%%%%%%%%%%%%%%%%%%%%%%%%%%%%%%%%%%%%%%%%%%%%%%%%%%%%%%%%%%

The Hamiltonian is also invariant under three fold rotations about the sites
and two fold rotations about the center of the links. We refer to the two-fold
rotation symmetry as inversion symmetry. At half filling, the system also has
particle-hole (chiral) symmetry, $c_i\rightarrow (-1)^{p_i}c^\dagger_i$, where
$p_i=0$ for $i$ belonging to one of the sublattices and $p_i=1$ for the other.

\section{\label{sec:III}Mean Field Approximation}
As mentioned earlier, we analyze the system using mean field theory.
The mean field decomposition we use is \cite{castro2013},
\begin{align}
\nonumber
n_in_j &\approx \left(\Delta_ic^\dagger_jc_j
+\Delta_jc^\dagger_ic_i\right)
-\chi_{\langle ij\rangle}c^\dagger_ic_j
-\chi_{\langle ij\rangle}^*c^\dagger_jc_i\\
&-\frac{1}{V}\left(\Delta_i^2+\Delta_j^2
-\vert\chi_{\langle ij\rangle}\vert^2\right),\\
\frac{1}{V}\chi_{\langle ij\rangle}&=\langle c^\dagger_jc_i\rangle_{MF},
~~~~~~
\frac{1}{V}\Delta_i=\sum_{j(i)}\langle c^\dagger_jc_j\rangle_{MF},
\label{sce}
\end{align}
%%%%%%%%%%%%%%%%%%%%%%%%%%%%%%%%%%%%%%%%%%%%%%%%%%%%%%%%%%%%%%%%%%%%%%%%%%%%%%%%%%%%%%%%%%%%%%%%%%%
where $j(i)$ denotes all the nearest neighbors of $i$. The self consistency
equations (\ref{sce}) have to be solved keeping the number density fixed.

The single particle mean field Hamiltonian in momentum space is the $6 \times
6$ matrix $h_{MF}(\vec k)=h_0(\vec k)+h_F(\vec k,\Delta,\chi)$ where $h_0$ is
the single particle non-interacting part of the Hamiltonian and $h_F$
represents the interaction with the mean field parameters; $\vec k$ takes values in
the reduced (magnetic) BZ. There are $6$ real charge density order
parameters represented by $\Delta_{(\alpha,a)}$ and $9$ complex bond order
parameters represented by $\chi_{\langle (\alpha,a)(\beta,b)\rangle}$ where
$\alpha,\beta$ label the original unit cells in the magnetic unit cell and $a,b$
label the two sublattices of the honeycomb lattice. Our mean field ansatz
allows the breaking of the translational, rotational and inversion
symmetries of the system.

We solve the self consistency equations, Eq.~\eqref{sce}, for $\Delta$ and
$\chi$ for both choices of the unit cell, {\bf I} and {\bf II}. These
correspond to different mean field ansatze.  The complex bond order parameters
$\chi_{\langle (\alpha,a)(\beta,b)\rangle}$ and real charge order parameters
$\Delta_{\alpha,a}$ are solved by an iterative method using the self
consistency equations, Eq.~\eqref{sce}, at a given filling and $V$. We
summarize the algorithm used here. (i) We start the iteration with a
random initial guess of $\chi_{\langle (\alpha,a)(\beta,b)\rangle}$ and
$\Delta_{\alpha,a}$, (ii) diagonalize $H_{MF}$ using $\chi_{\langle
(\alpha,a)(\beta,b)\rangle}$ and $\Delta_{\alpha,a}$, (iii) Tune 
the chemical potential by fixing the number of particles (iv) calculate the
expectation value of the link operators and the number
operators on each site in the magnetic unit cell. Using this we compute
the new values of 
$\chi_{\langle (\alpha,a)(\beta,b)\rangle}$ and $\Delta_{\alpha,a}$ from
Eqs.~\eqref{sce}. The whole process from step (ii) to (iv) is repeated
until all the quantities converge.  We repeat this process for various initial
guesses and often find different mean field solutions. Comparing the energies
of these solutions, we pick up the lowest energy state as the ground state of
the interacting Hamiltonian. This method is repeated for $V\in(1,10)$ for
$m=1,2,3$ and we get various complex phases which are discussed in the
following section.

\section{\label{sec:IV}Results}
\begin{figure}[hbtp]
\begin{center}
\includegraphics[width=0.38\textwidth]{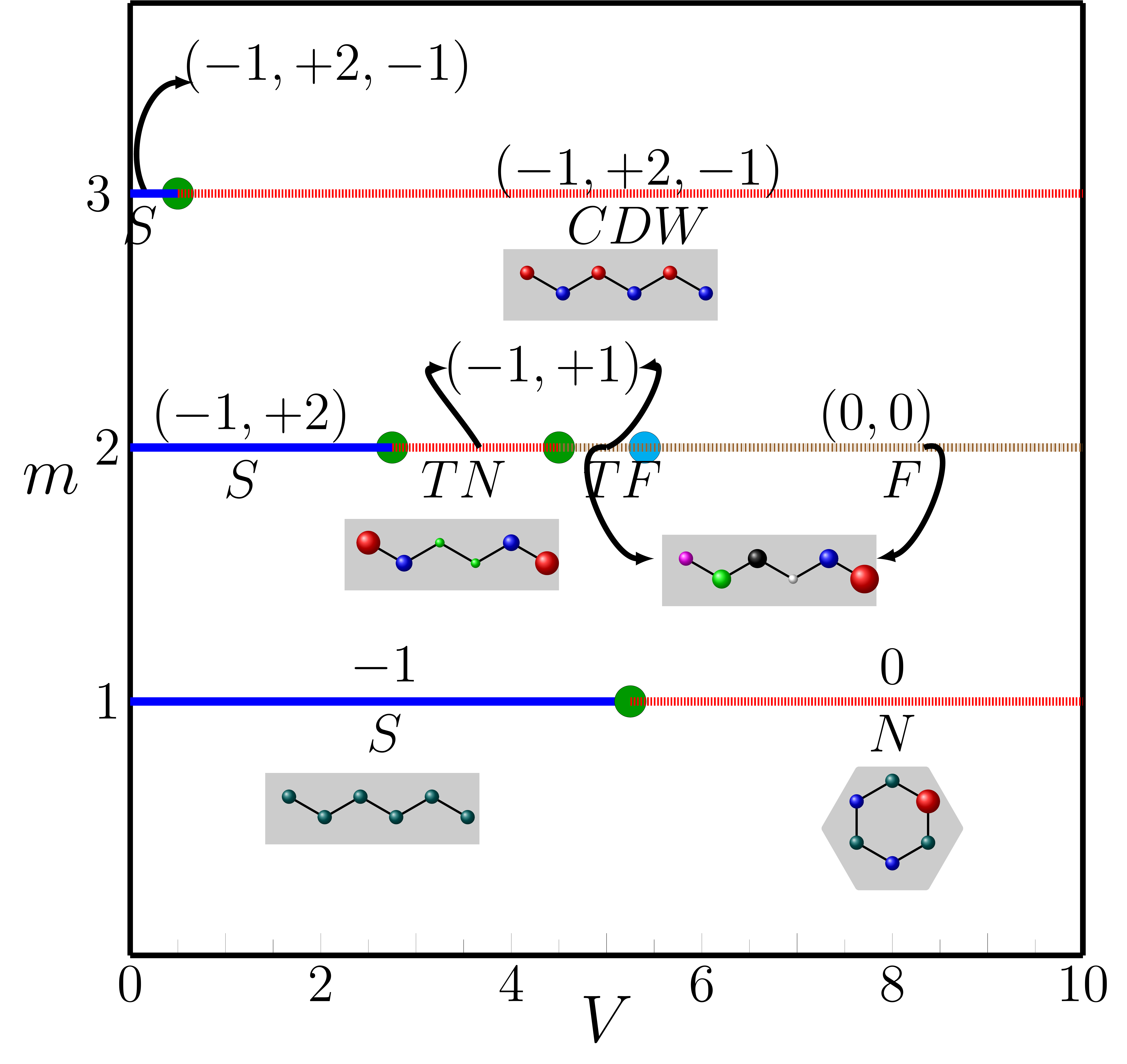}\\[0.5cm]
\caption{\label{fig2}(Color online) Phase diagram for $m=1,2,3$ filling
as $V$ is varied. The charge distribution in each phase is
shown by the unit cell whose repetition gives the full lattice picture of
the ground state. The Chern number distribution of the filled bands is given in
bracket. For example in the $m=2$ case, the left number in bracket is for the lowest band
and the next number is for the second band. The green circle is the transition point.
The blue circle is the point where the Chern numbers of the filled bands change without
any change in the Hall conductivity.}
\end{center}
\end{figure}

This section describes various phases which  are the minimum energy solutions of the self consistency equations.
For small $V$, at all three fillings, the system is in a phase 
where the mean field ground state is the same as that of the non-interacting case.
We denote this as the symmetric phase ($S$). The charge density is uniform and and all the
symmetries of the Hamiltonian are preserved in this phase. At larger $V$, we find several
charge ordered phases that break translational and rotational symmetries of the system. 
We characterize the charge ordering by the dipole moment
($P^\mu$) and the quadrupole moment ($Q^{\mu\nu}$) of the single particle density.
These are defined as, 
\begin{align}
&P^\mu\equiv\frac{1}{N}\sum_i~R^\mu_i\Delta_i,\\
&Q^{\mu\nu}\equiv\frac{1}{N}\sum_i\left(2R^{\mu}_iR^\nu_i
-\delta^{\mu\nu}\bm{R_i}\cdot\bm{R_i}\right)\Delta_i~,
\end{align}
where $R^\mu_i,~(\mu=1,2)$ are the components of the position vector $\bm{R_i}$
at site $i$ and $N$ is the total number of original unit cells in
the lattice. It is easy to show that $Q^{\mu\nu}$ can be non-zero only if the
rotational symmetry is broken and $P^\mu$ can be non-zero only if both
rotational and inversion symmetries are broken. The phases with broken symmetry,
their charge ordering and Hall conductivities are discussed below and depicted 
in Fig.~\ref{fig2}. Here, the charge distribution is shown in a magnetic unit
cell. The sites with same charge are shown in same color and size spheres. The charge
 greater in magnitude is represented with a bigger sphere. The Chern  number
distribution for all the filled  bands  in each filling is given in the bracket.
The left most number in the bracket represents the Chern number 
for the lowest band and next number
is for the next filled band. The right most number in the bracket gives  the 
Chern number of the highest band filled.

At half filling ($m=3$, Dirac regime), for both choices of unit cells,
there is a continuous transition from the symmetric phase to a charge density
wave (CDW) phase at $V=0.45$. This phase preserves all but the inversion
symmetry of the system. The charge distribution in the unit cell is shown in
Fig.~\ref{fig2}. The dipole moment and quadrupole moment remain zero in this
phase. As $V$ is increased, the CDW strengthens but there are no other
transitions for $V\le 10$. The Hall conductivity remains zero and the Chern
number distribution in the bands remains unaltered and is similar to the $S$ phase. 
%%%%%%%%%%%%%%%%%%%%%%%%%%%%%%%%%%%%%%%%%%%%%%%%%%%%%%%%%%%%%%%%%%%%%%%%%%%%%%%%%%%%%%%%%%%%%%%%%%%%%%%%%%%%%%%%
%%%%%%%%%%%%%%%%%%%%%%%%%%%%%%%%%%%%%%%%%%%%%%%%%%%%%%%%%%%%%%%%%%%%%%%%%%%%%%%%%%%%%%%%%%%%%%%%%%%%%%%%%%%%%%%%%

\begin{figure}[hbtp] \begin{center}
\includegraphics[scale=0.075]{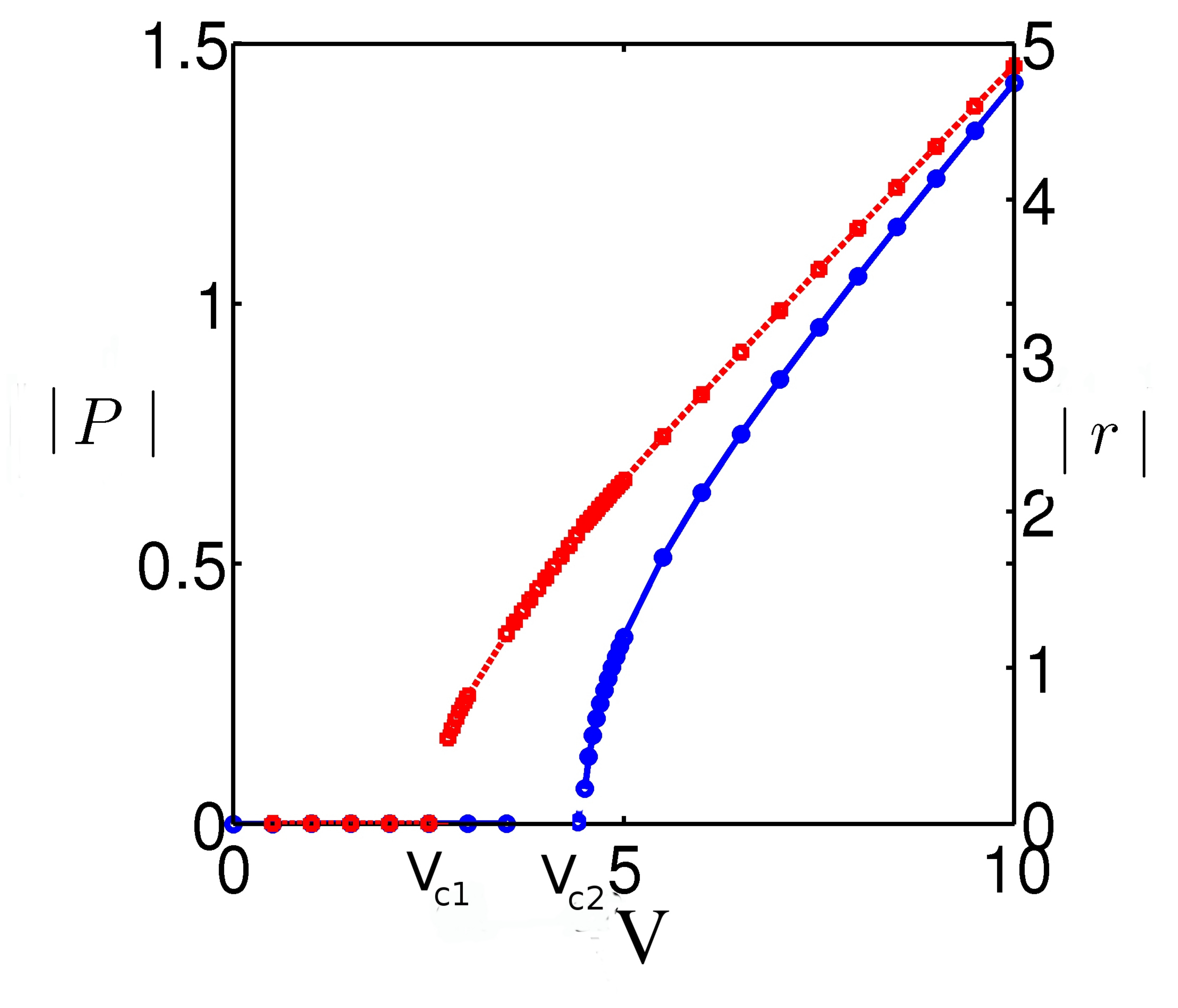} 
\caption{\label{qmdm}(Color online) The quadrupole and dipole moments
  as a function of $V$ in the three phases for $m=2$. The blue line
  represents the polarization (i.e. dipole moment per original unit
  cell)and the red line represents the magnitude of the eigenvalue of
  the quadrupole moment matrix per original unit cell, $|r|$.}
\end{center}
\end{figure}

For the van Hove filling, $m=2$, the energy of the
mean field solution for the unit cell {\bf I} is always lower in energy than that of {\bf II}
for the range of $V$ considered here.
At $V=V_{c1}=2.744$, there is a first order
transition from the symmetric phase to a charge ordered nematic phase where translational
and three-fold rotational symmetries are broken but the inversion symmetry of the system is
intact about the center of the $A_2B_2$ links. It has non-zero quadrupole moment but 
the dipole moment is zero as shown in Fig.~\ref{qmdm}.
Quadrupole moment is a $2\times2$ traceless matrix. In Fig.~\ref{qmdm}, we plot the magnitude of
the eigenvalue of the quadrupole moment denoted as $|r|$.
The blue line represents the plot for the magnitude of the dipole moment per original unit cell 
varying as a function of $V$. This
Landau transition is accompanied by a topological transition where the Hall
conductivity changes from $e^2/h$ to zero. However, the Chern numbers of the
two occupied bands individually are non-zero. The Chern number of the lowest band is $-1$ and that
of the second band is $1$ (Fig.~\ref{fig2}), hence we name the phase as the topological nematic phase ($TN$). 
%%%%%%%%%%%%%%%%%%%%%%%%%%%%%%%%%%%%%%%%%%%%%%%%%%%%%%%%%%%%%%%%%%%%%%%%%%%%%%%%%%%%%%%%%%%%%%%%%%%%%%%%%%%%%%%
%%%%%%%%%%%%%%%%%%%%%%%%%%%%%%%%%%%%%%%%%%%%%%%%%%%%%%%%%%%%%%%%%%%%%%%%%%%%%%%%%%%%%%%%%%%%%%%%%%%%%%%%%%%%%%%%

As $V$ is increased, we encounter another first order transition at
$V=V_{c2}=4.5$ to a phase where the inversion symmetry is also broken. The
system now develops non-zero dipole moment in addition to the quadrupole moment as shown in Fig.~\ref{qmdm}.
The Chern number distribution in the occupied bands remains the same as that in the
$TN$ phase  (Fig.~\ref{fig2}). We call this phase, the topological ferri-electric phase ($TF$). 
The two occupied bands touch at $V=5.41$ and there is a redistribution of Chern numbers between these bands.
For $V>5.41$ the Chern number of each of the filled bands
become zero (Fig.~\ref{fig2}). This phase is denoted as the ferri-electric phase ($F$).

In the Fermi regime, $m=1$, the energy for the mean field solution with unit cell
choice {\bf II} is lower than that of {\bf I}. Here, we find a first order
transition from the symmetric phase to a nematic phase ($N$) (Fig.~\ref{fig2}),
similar to that described for $m=2$, at $V=5.265$. The charge distribution has no
dipole moment but has non-zero quadrupole moment. The Hall conductivity
changes from $\sigma_H=-e^2/h$ to zero at this transition. Thus, this is also
a first order Landau transition accompanied by a topological transition.

The charge ordered phases with broken translational and three-fold rotational symmetries also have
anisotropic bond order parameters, $\chi_{\langle(\alpha,a)(\beta,b)\rangle}$. 
The anisotropic magnitudes of the order parameters are shown in Fig.~\ref{fig4}
for $m=1~{\rm and}~2$. For $m=2$, the bond order parameters also
acquire anisotropic phases which manifest as circulating currents.
%%%%%%%%%%%%%%%%%%%%%%%%%%%%%%%%%%%%%%%%%%%%%%%%%%%%%%%%%%%%%%%%%%%%%%%%%%%%%%%%%%%%%%%%%%%%%%%%%%%%%%%%%%%%%%%%%
\begin{figure}[h!]
\begin{center}
\includegraphics[scale=0.12]{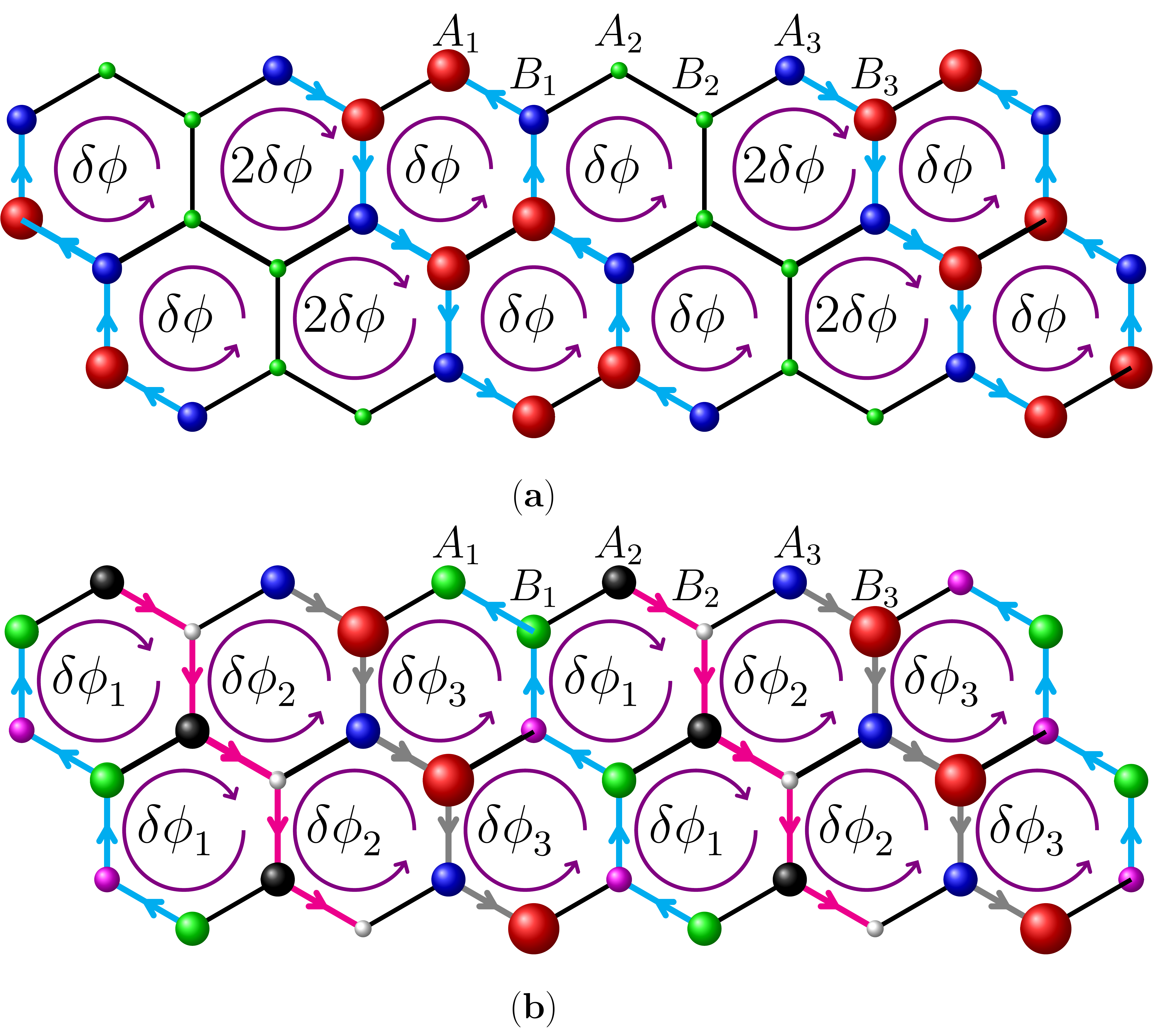}
\caption{\label{currents}(Color online) Flux in the plaquette and
  current flowing on the links on the lattice for $m=2$ in (a) nematic
  phase (b) ferri-electric phase. The flux distribution is shown after
  subtracting out the background flux $2\pi/3$. In ferri-electric
  phase, $\delta\phi_1=\delta\phi_3+\delta\phi_2$.  The bonds with
  arrows are the links on which there is non-zero current. Same color
  represents same magnitude of current.}
\end{center}
\end{figure}
%%%%%%%%%%%%%%%%%%%%%%%%%%%%%%%%%%%%%%%%%%%%%%%%%%%%%%%%%%%%%%%%%%%%%%%%%%%%%%%%%%%%%%%%%%%%%%%%%%%%%%%%%%%%%%%%%

There is no current on the links in the symmetric phase. 
Fig.~\ref{currents} shows the current on the links on the lattice in
nematic and ferri-electric phases. In nematic phase, currents of equal magnitude
flow on the links $A_1B_1$ and $A_3B_3$ (links with arrows) in opposite direction 
(as shown in Fig.~\ref{currents}a) such that the
inversion symmetry (about $A_2B_2$ link) of the system is preserved.
There is no current on $A_2 B_2$ links. In ferri-electric phase, the current
flows on the links $A_1B_1$, $A_2B_2$ and $A_3B_3$
(Fig.~\ref{currents}b) such that the total current on the links in the
magnetic unit cell is zero. The current distribution shows that the inversion
symmetry of the system along with the rotational symmetry is
broken. 
This figure also gives the distribution of flux in the
plaquettes of the system in nematic and ferri-electric phases. In
Fig.~\ref{currents} the flux distribution is shown after subtracting
out the background flux $2\pi/3$. As seen in Fig.~\ref{currents}a, the
staggered flux in the nematic phase is distributed in a way that the
inversion symmetry of the lattice is preserved unlike that in the
ferri-electric phase where the flux distribution shows inversion
symmetry breaking. In both the cases the total flux in the magnetic
unit cell is zero.

We now compare these staggered flux patterns in the lattice with those
discussed by Castro et al. \cite{castro2011}, where they analyzed the system without a
magnetic field. 
In their paper, they have described about two interaction induced spontaneously time reversal symmetry (TRS)
broken  phases, {\bf T-I} and {\bf T-II} phases. While in  {\bf T-II} phase the  inversion
symmetry is preserved, {\bf T-I} phase breaks the inversion symmetry.
We see that the flux pattern for nematic phase is similar to
that of their {\bf T-II} phase. In both
these cases inversion symmetry of the system is preserved. The staggered flux
pattern in ferri-electric phase shows inversion symmetry breaking in the system
as is the case for {\bf T-I} phase in
\cite{castro2011}; However, in ferri-electric phase every plaquette has
non-zero flux passing through it and the staggered flux in all the three
plaquettes forming the magnetic unit cell have different magnitude  unlike the
{\bf T-I} phase. The current on the links in our system form a stripe pattern
unlike the case without the magnetic field \cite{castro2011} where the  current
on the links form closed loops.

To try and get some insight to the mechanism of the topological transitions 
accompanying the Landau transitions, we examine the pattern of the anisotropic
magnitudes of the bond order parameters shown in Fig.~\ref{fig4}. The bond
strength distribution for the $TN,~TF~{\rm and}~F$ phases indicate that the
mean-field Hamiltonian resembles that of weakly coupled ribbons, namely a
quasi-$1d$ system. The coupling becomes weaker as the interaction strength is
increased. In the limit of completely decoupled ribbons, Chern numbers for all
the bands can be equal to zero and this happens for $V>5.41$.  Thus it seems
that the anisotropy of the bond order parameters drives the change of Chern
numbers. The $N$ phase at $m=1$ is similar except that it tends to a system of
weakly coupled clusters as shown in Fig.~\ref{fig4}b.

However, while we believe that there is some truth to the argument given above,
there are some caveats. Firstly, the transitions need not happen at a finite
value of $V$ but could happen only at $V=\infty$. Secondly when the system gets
decoupled, the Chern numbers of the individual bands can get ill-defined due to
degeneracies. Both these issues can be illustrated in the half-filled case of
$m=3$. As shown in Fig.~\ref{fig2}, the Chern numbers in this case remain
unchanged throughout the range, $0\le V\le 10$. We now show that this is true
for arbitrary $V$. As mentioned earlier, the single particle mean field
Hamiltonian is of the form, $h_{MF}(\vec k)=h_0(\vec k)+h_F(\vec k,\chi,\Delta)$
For the half filled case, the system is isotropic and the bond order parameters
simply scales $h_0(\vec k)$ and the charge modulation is $\pm\Delta$ for the
two sublattices. Since $h_0(\vec k)$ does not couple the two sublattices, we 
can write $\vec h_{MF}(\vec k)$ as,
\begin{equation}
\label{mfhcdw}
h_{MF}(\vec k)=\left(\begin{array}{cc}\Delta&F(\vec k)\\
F^\dagger(\vec k)&-\Delta\end{array}\right)
\end{equation}
It can be proved, as we do in Appendix \ref{sec:A1}, that the Chern
numbers of the above Hamiltonian are independent of $\Delta$. As 
$V\rightarrow\infty$ we also have $\Delta\rightarrow\infty$. In 
the limit, there are two degenerate ground state corresponding
to all the particles occupying one of the sublattices. However, all the
the negative energy single particle states are degenerate and thus
the three lower bands are completely degenerate. Thus the individual
Chern numbers are ill-defined at $V=\infty$. At $\Delta=0$ (the symmetric
phase), the middle band Chern numbers are ill-defined due to the Dirac
points. However, the system is particle-hole symmetric in this phase and 
the Chern numbers get fixed by demanding that the sum of the Chern
numbers, which is proportional to the Hall conductivity, is zero. Thus
the Chern numbers in the half filled case are unchanged in the range
$0\le V<\infty$.

%%%%%%%%%%%%%%%%%%%%%%%%%%%%%%%%%%%%%%%%%%%%%%%%%%%%%%%%%%%%%%%%%%%%%%%%%%%%%%%%%%%%%%%%%%%%%%%%%%%%%%%%%%%%%%%%
\begin{figure}[hbtp] \begin{center}
\includegraphics[width=0.22\textwidth]{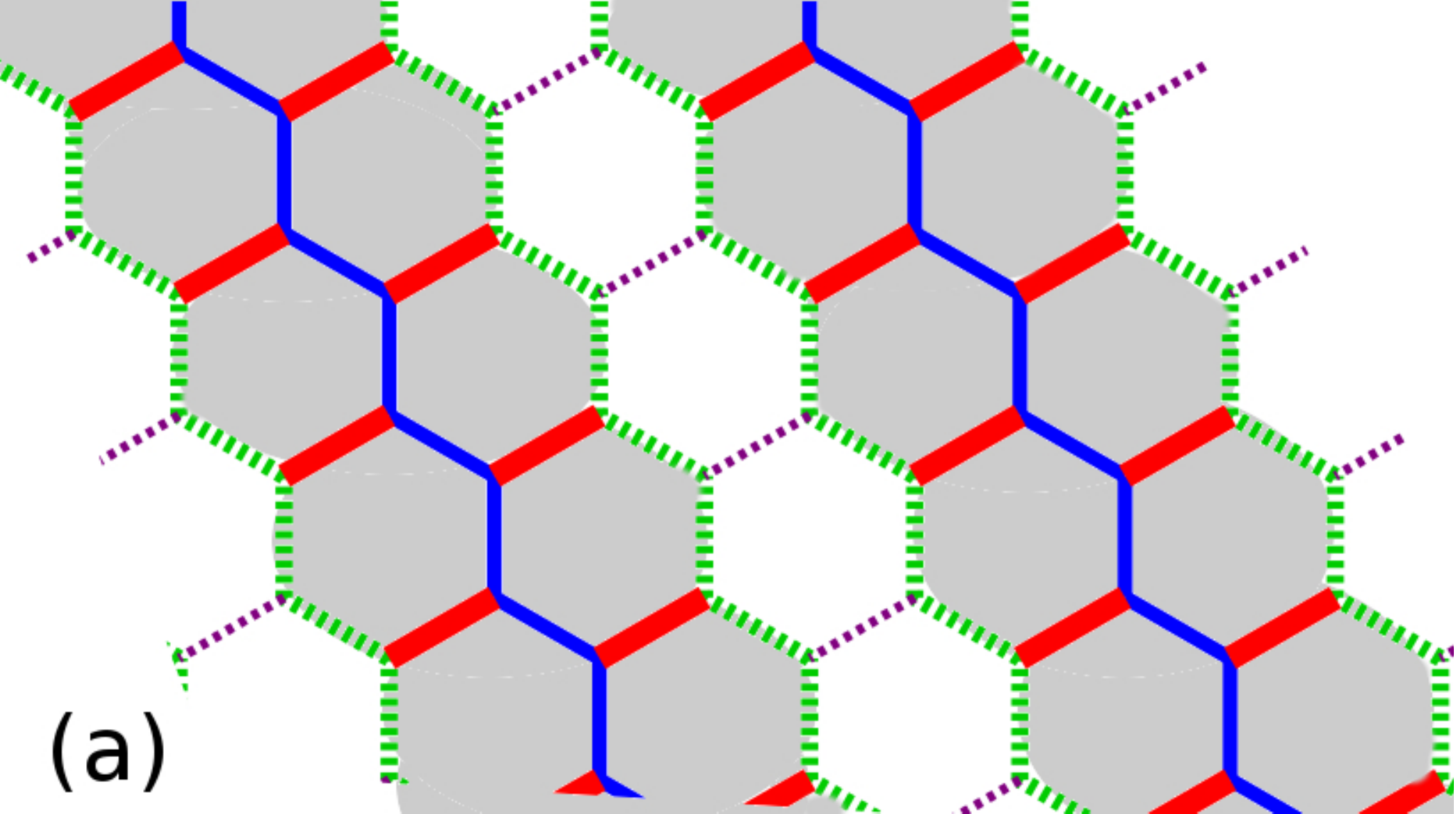}
\includegraphics[width=0.22\textwidth]{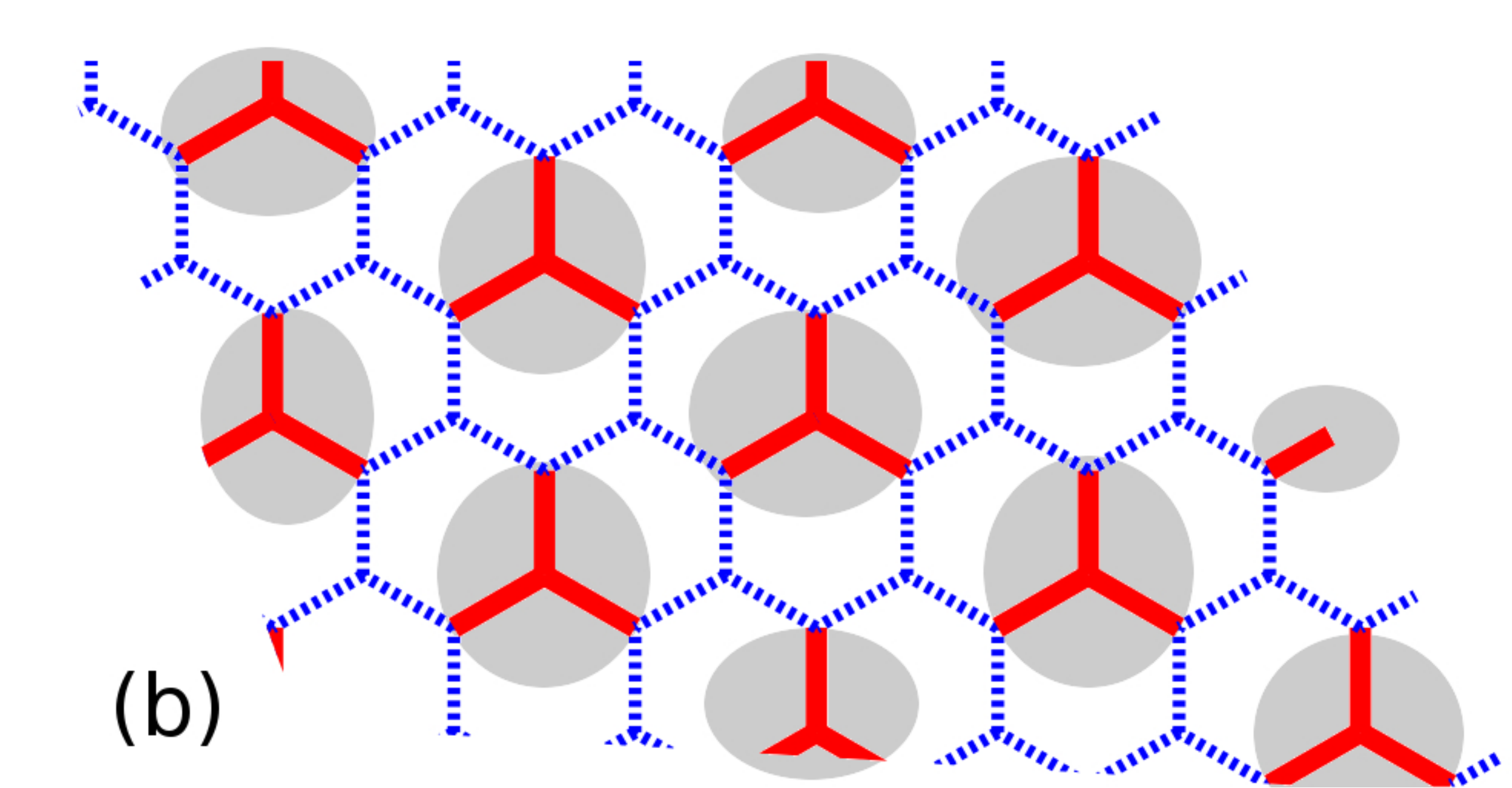} 
\end{center}
\caption{\label{fig4}(Color
online) (a) Bond order parameter strengths for $m=2$ broken symmetry phases.
The red bonds are the strongest and the dotted violet bonds the weakest.
(b) Bond order parameters for the $m=1$ broken symmetry phase. The red bonds are
stronger than the dotted blue bonds.}
\end{figure}
%%%%%%%%%%%%%%%%%%%%%%%%%%%%%%%%%%%%%%%%%%%%%%%%%%%%%%%%%%%%%%%%%%%%%%%%%%%%%%%%%%%%%%%%%%%%%%%%%%%%%%%%%%%%%%%%%%

\section{\label{sec:V}Geometry of the ground states}
The structure of incompressible liquids is described by the pair correlation
function. For our system it is defined as,
\begin{equation}
\Gamma_{\alpha a,\beta b}\left(\bm{R}_I-\bm{R}_J\right)=
\langle c^\dagger_{Ia\alpha}c^\dagger_{Jb\beta}c_{Jb\beta}c_{Ia\alpha}\rangle
\end{equation}
where $I,J$ represent the position of the magnetic unit cell.
The anisotropic state of the system can be understood in terms of 
this pair correlation function. The nematic
order parameter has been related to the quantum metric \cite{maciejko2013}. 
In this section, we would therefore like to study the pair correlation function in the light of	quantum
geometry described by the quantum metric.
Resta \cite{resta2006} has derived an exact relationship between the 
momentum space quantum metric 
\cite{resta2011} averaged over the BZ, $\bar g^{\mu\nu}$, and the pair 
correlation function, more precisely, to the structure factor\\
\begin{multline}
\bar g^{\mu\nu}=\frac{1}{L^2}\sum_{I\alpha a,J \beta b}
\left(R^\mu_{I\alpha a}-R^\mu_{J\beta b}\right)
\left(R^\nu_{I\alpha a}-R^\nu_{J\beta b}\right)\\
\label{gmoment}
\qquad S_{\alpha a,\beta b}( \bm{R}_I-\bm{R}_J)~,
\end{multline}
where $L^2$ is the area of the system and
$S_{\alpha a,\beta b}(\bm{R}_I-\bm{R}_J)=
\langle c^\dagger_{I\alpha a}c_{I\alpha a}\rangle
\langle c^\dagger_{J\beta b}c_{J\beta b}\rangle
-\Gamma_{\alpha a,\beta b}\left(\bm{R}_I-\bm{R}_J\right)$. 

Thus $\bar g^{\mu\nu}$ is the second moment of the 
structure factor. It therefore characterizes the shape of the pair
correlation function just as the real space metric
introduced by Haldane \cite{haldane2011} for homogeneous quantum Hall
systems. To obtain a more precise relation between the two, we need to 
examine the weak field (large $q$) limit. We postpone this for future work. 
\begin{figure}[h!] \begin{center}
\includegraphics[scale=0.08]{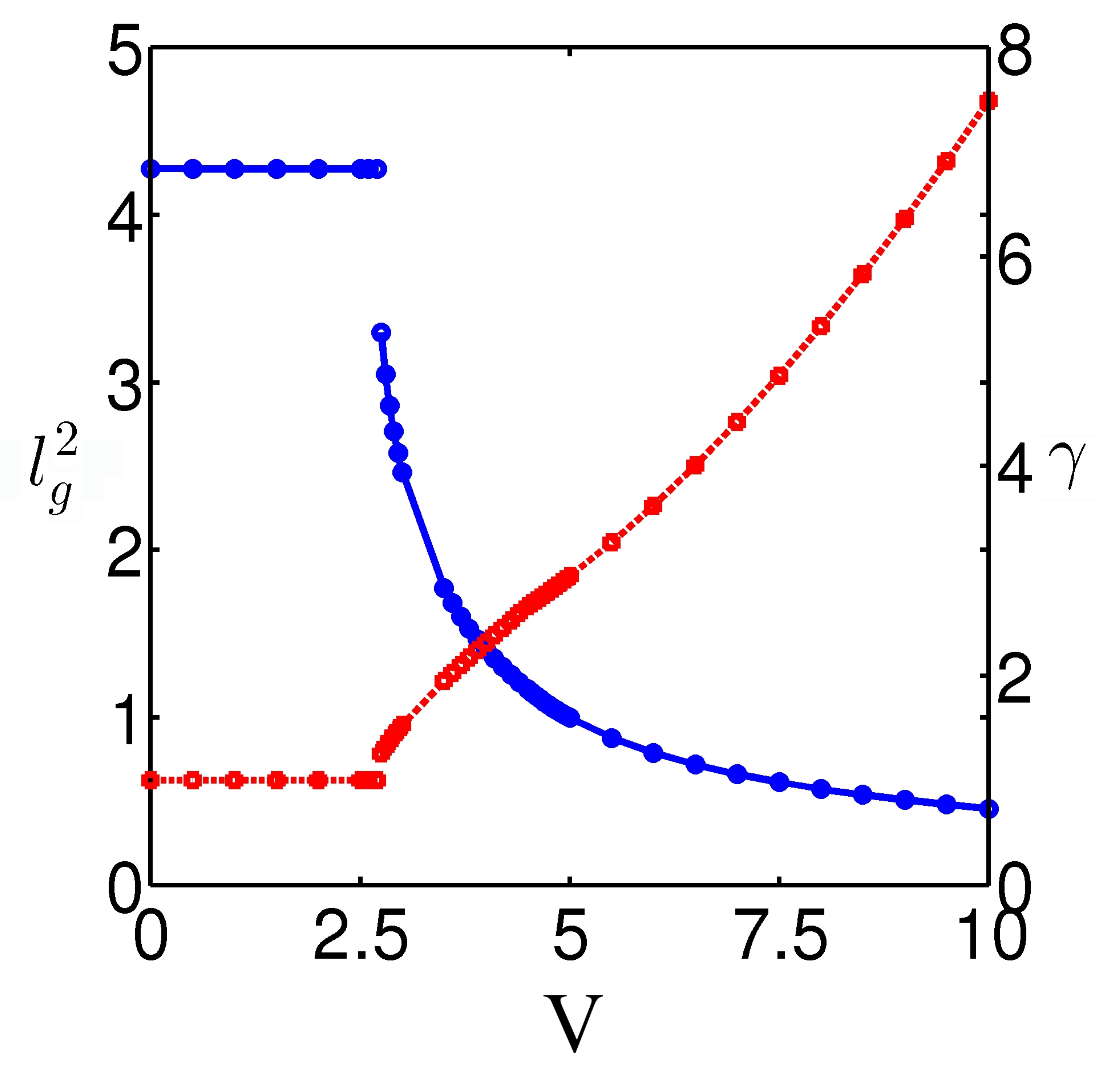}
\caption{\label{metrics}(Color
online) $l^2_g=\sqrt{g_1g_2}$ and $\gamma$ vs $V$ for $m=2$.
The blue line represents $l_g$ and
the red line represents $\gamma$. The discontinuity in the plot shows the first order nature
of the phase transition.}
\end{center}
\end{figure}

The metric is a second rank symmetric tensor and has three independent
components. These can be taken to be the orientation of the principle axis,
$\theta$ and the two eigenvalues $g_1,g_2$. Eq.~\eqref{gmoment} shows that
the spread of the pair correlation function in the directions along and normal
to the principal axis is given by $\sqrt{g_1}$ and $\sqrt{g_2}$. Therefore, the areal
extent of the correlations $\sim \sqrt{g_1g_2}\equiv l_g^2$, the 
square root of the determinant of the $\bar g^{\mu\nu}$. The ratio of the two
eigenvalues $\gamma=g_1/g_2$ is a measure of its anisotropy. $l_g^2$ and 
$\gamma$ are plotted as a function of the interaction strength in 
Fig.~\ref{metrics}. $l^2_g$ decreases while the anisotropic parameter $\gamma$ increases on increasing $V$.

\begin{figure}[h!] \begin{center}
\subfloat[]{\label{pc1}\includegraphics[scale=0.07]{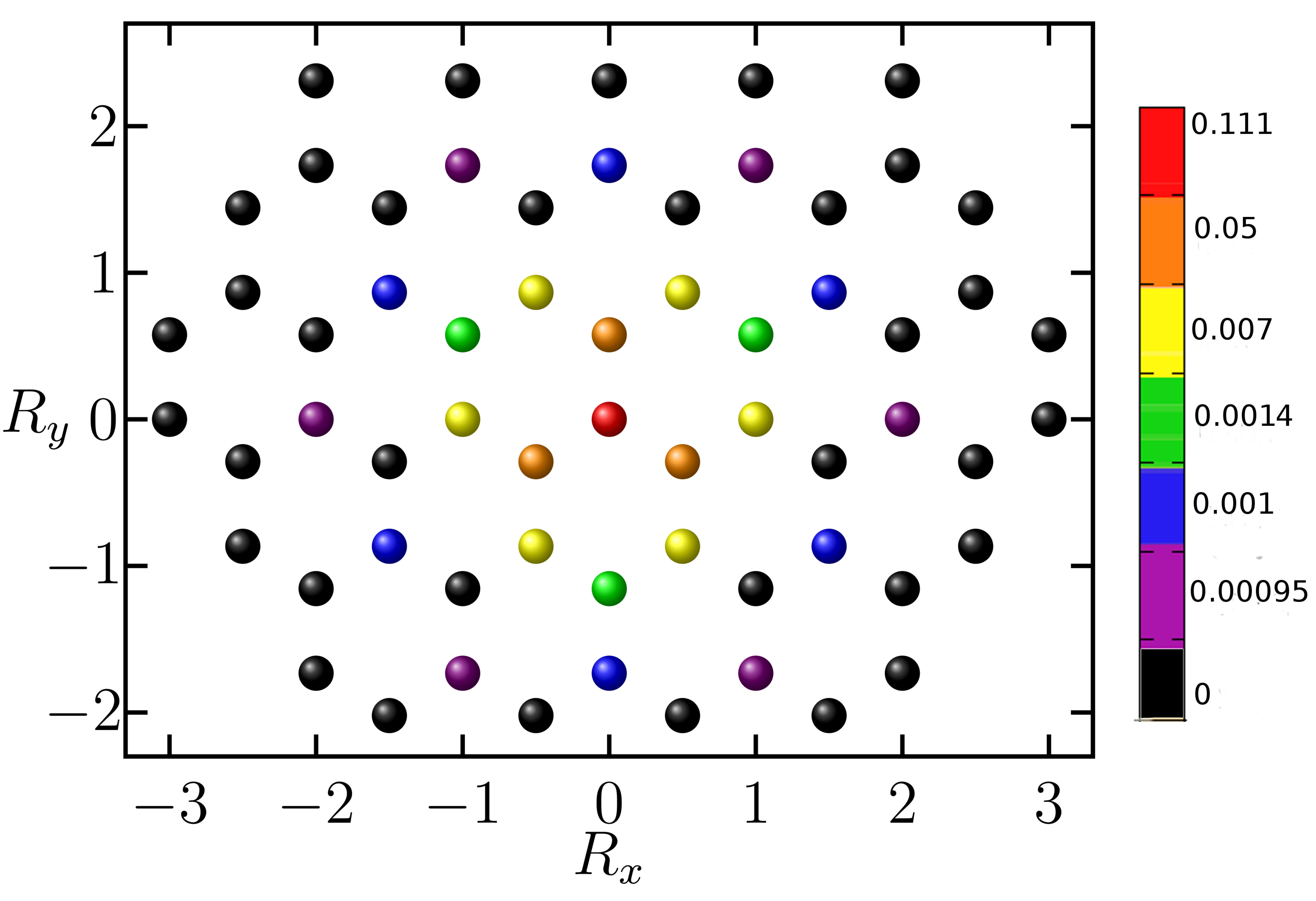}}\\
\subfloat[]{\label{pc2}\includegraphics[scale=0.07]{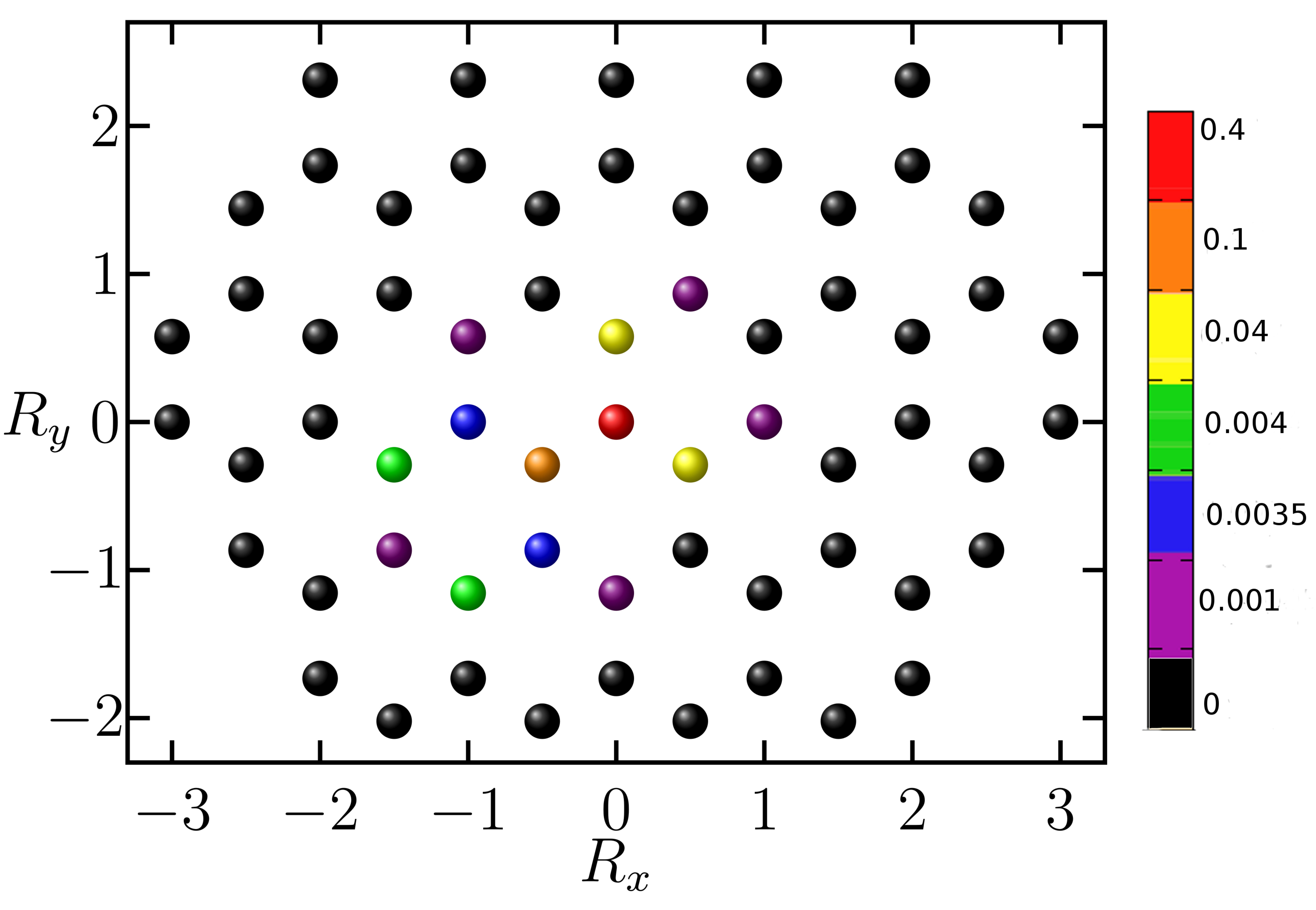}}
\caption{\label{pc}(Color online) Pair correlation function on lattice
  in (a) symmetric phase showing $2\pi/3$ rotation preserved, (b)
  nematic phase for $m=2$ showing $2\pi/3$ rotation broken.}
\end{center}
\end{figure}

The structure function reflects the symmetry (or the lack of it) of the ground
state. Fig. \ref{pc1} shows the structure function plotted in the real
space for symmetric phase for $m=2$.  $R_x$ and $R_y$ give the
position of the lattices in the real space in Cartesian coordinates.
We see that in this plot, the pair correlation function is invariant
under $2\pi/3$ rotation since the symmetric phase preserves the
rotational symmetry of the system.
Fig. \ref{pc2} shows the structure function plotted in the real
space for nematic phase for $m=2$.  In this plot, the structure
function is not invariant under $2\pi/3$ rotation since in nematic
phase the rotational symmetry is broken.
%%%%%%%%%%%%%%%%%%%%%%%%%%%%%%%%%%%%%%%%%%%%%%%%%%%%%%%%%%%%%%%%%%%%%%%%%%%%%%%%%%%%%%%%%%%%%%%%%%%%%%%%%%%%%%
%%%%%%%%%%%%%%%%%%%%%%%%%%%%%%%%%%%%%%%%%%%%%%%%%%%%%%%%%%%%%%%%%%%%%%%%%%%%%%%%%%%%%%%%%%%%%%%%%%%%%%%%%%%%%%%

The average metric is a multiple of identity in the
symmetric phase. In the nematic and ferri-electric phases it becomes
anisotropic, with its principle axis aligned with one of the basis vectors.

\section{\label{sec:VI}Conclusion}

To summarize our results, we show that the nearest neighbor repulsive
interaction induces a charge ordering, as we intuitively expect. At strong
interactions, translation symmetry broken, anisotropic charge distributions
become energetically favorable. The anisotropy in the particle density can be
characterized by the quadrupole and dipole moments. The anisotropy and the
spatial extent of the pair correlations are characterized by the quantum metric
averaged over the BZ. 

The first order transition for $m=1,2$ from the symmetric to the nematic phase
is accompanied with change of topology which is reflected in the change in the
Hall conductivity. Though the Hall conductivity is zero in the nematic phase,
in $m=2$ band, the filled bands individually have non-trivial topology with
non-zero Chern number. Some insight for the mechanism of this transition comes
from examining the magnitudes of the bond order parameters,
$\chi_{\langle\alpha a,\beta b\rangle}$. The 2d lattice looks like a set of weakly coupled 1d
ribbons or clusters. In the extreme limit of decoupled ribbons, the Chern
number is zero and hence will remain so for weak coupling as well. So a Chern
number change accompanies the first order nematic transition when the weakening
is significant. 

In conclusion, our results show that interactions induce interesting and
complex phases in the Hofstadter regime of the honeycomb lattice. 

The fractal structure is understood as arising from the interplay between the 
two independent length scales, the periodicity and the magnetic length. We have
shown that the interaction induces charge ordering which breaks the 
translational symmetry and thus changes the periodicity.
The effect of this change
of the periodicity on the fractal structure of the Hofstadter butterfly requires a study of the translation
symmetry breaking pattern for other values of $\phi/\phi_0$ as well. This
work is in progress.

\acknowledgments  We are grateful to A. M. M. Pruisken, Dibyakrupa Sahoo,
Jainendra K. Jain, Mukul S. Laad, S. Arya and Vinu Lukose
for useful discussions.

\appendix
\section{Chern numbers at half filling}\label{sec:A1}

The single particle mean field Hamiltonian for the half filled case
describing the CDW state is given in equation (\ref{mfhcdw}). The
eigenvalue equation can be written as,
\begin{equation}
\left(\begin{array}{cc}\Delta&F(\vec k)\\F^\dagger(\vec k)&-\Delta
\end{array}\right)
\left(\begin{array}{c}\psi_A(\vec k)\\\psi_B(\vec k)\end{array}\right)=
E(\vec k)
\left(\begin{array}{c}\psi_A(\vec k)\\\psi_B(\vec k)\end{array}\right)
\end{equation}
where $\psi_{A(B)}(\vec k)$ are $q$-component column vectors. 
They can be constructed in terms of the spectrum of the positive,
semi-definite, hermitian matrix,
$F(\vec k)F^\dagger(\vec k)$. We denote,
\begin{equation}
F(\vec k)F^\dagger(\vec k)\chi^n(\vec k)=\epsilon_n^2(\vec k)\chi^n(\vec k)
\end{equation}
where, $n=1,\dots,q$ and we choose $\chi^n$ to be ortho-normalized. 
The above equation implies that the eigenvalues of 
$F^\dagger(\vec k)F(\vec k)$ are the same, since,
\begin{equation}
F^\dagger(\vec k)F(\vec k)\left(F^\dagger(\vec k)\chi^n(\vec k)\right)
=\epsilon_n^2(\vec k)\left(F^\dagger(\vec k)\chi^n(\vec k)\right)
\end{equation}
 
Further, the inversion (two fold rotation) transformation
relates $F(\vec k)$ to $F^\dagger(-\vec k)$, 
\begin{equation}
{\cal I}F(\vec k){\cal I}^\dagger=F^\dagger(-\vec k)
\end{equation}
where,
\begin{equation}
{\cal I}=\left(\begin{array}{ccc}0&0&1\\0&1&0\\1&0&0\end{array}\right)
\end{equation}
consequently, we have $\epsilon_n(\vec k)=\epsilon_n(-\vec k)$ and
\begin{equation}
F^\dagger(\vec k)F(\vec k)\left({\cal I}\chi^n(-\vec k)\right)
=\epsilon_n^2(\vec k)\left({\cal I}\chi^n(-\vec k)\right)
\end{equation}

The eigenvectors of $h_{MF}(\vec k)$, $\psi^{\pm n}(\vec k)$
corresponding to the eigenvalues $E_{\pm n}(\vec k)=
\pm\sqrt{\epsilon_n^2(\vec k)+\Delta^2}$ are given by,
\begin{eqnarray}
\psi^{+n}(\vec k)&=&\left(\begin{array}{c}
\cos\frac{\theta_n(\vec k)}{2}~\chi^n(\vec k)\\
\sin\frac{\theta_n(\vec k)}{2}~{\cal I}\chi^n(-\vec k)\end{array}\right)
\\
\psi^{-n}(\vec k)&=&\left(\begin{array}{c}
-\sin\frac{\theta_n(\vec k)}{2}~\chi_n(\vec k)\\
\cos\frac{\theta_n(\vec k)}{2}~{\cal I}\chi^n(-\vec k)\end{array}\right)
\end{eqnarray}
where,
\begin{equation}
\cos\theta_n(\vec k)=\frac{\Delta}{\sqrt{\epsilon_n^2(\vec k)+\Delta^2}},~~
\sin\theta_n(\vec k)=
\frac{\epsilon_n(\vec k)}{\sqrt{\epsilon_n^2(\vec k)+\Delta^2}}
\end{equation}
The Pancharatnam-Berry curvatures of the negative energy bands (occupied
at half filling) are given by ${\cal B}^{-n}(\vec k)=
\epsilon_{ij}\partial_i{\psi^{-n}}^\dagger(\vec k)\partial_j\psi^{-n}(\vec k)$.
The Chern numbers are given by,
\begin{eqnarray}
\nonumber
\nu^{n}&=&\int_k\left(\epsilon_{ij}
\sin\theta_n\partial_i\theta_n\left(\tilde{\cal A}^n_j(\vec k)
+\tilde{\cal A}^n_j(-\vec k)\right)+\right.\\
&&\left.\epsilon_{ij}\left(\sin^2\frac{\theta_n}{2}\tilde{\cal B}^n(\vec k)
+\cos^2\frac{\theta_n}{2}\tilde{\cal B}^n(-\vec k)\right)\right)
\end{eqnarray}
where the integral is over the reduced BZ, 
$\tilde{\cal A}^n_i(\vec k)\equiv -i{\chi^n}^\dagger(\vec k)
\partial_i\chi^n(\vec k)$ and $\tilde{\cal B}^n(\vec k)=
\epsilon_{ij}\partial_i\tilde{\cal A}^n(\vec k)$. Since,
$\tilde{\cal A}^n_i(\vec k)=-\tilde{\cal A}^n_i(-\vec k),~
\tilde{\cal B}^n(\vec k)=\tilde{\cal B}^n(-\vec k)$ and
$\theta_n(\vec k)=\theta_n(-\vec k)$, we get our final result,
\begin{equation}
\nu^n=\int_k\tilde{\cal B}^n(\vec k)
\end{equation}
Thus, since $\chi^n(\vec k)$ is independent of $\Delta$, so are
the Chern numbers.

\section{Energy Band Diagram}\label{sec:A2}
\begin{figure}[!ht]
\begin{center}
\subfloat[]{\label{q3_s}\includegraphics[scale=0.072]{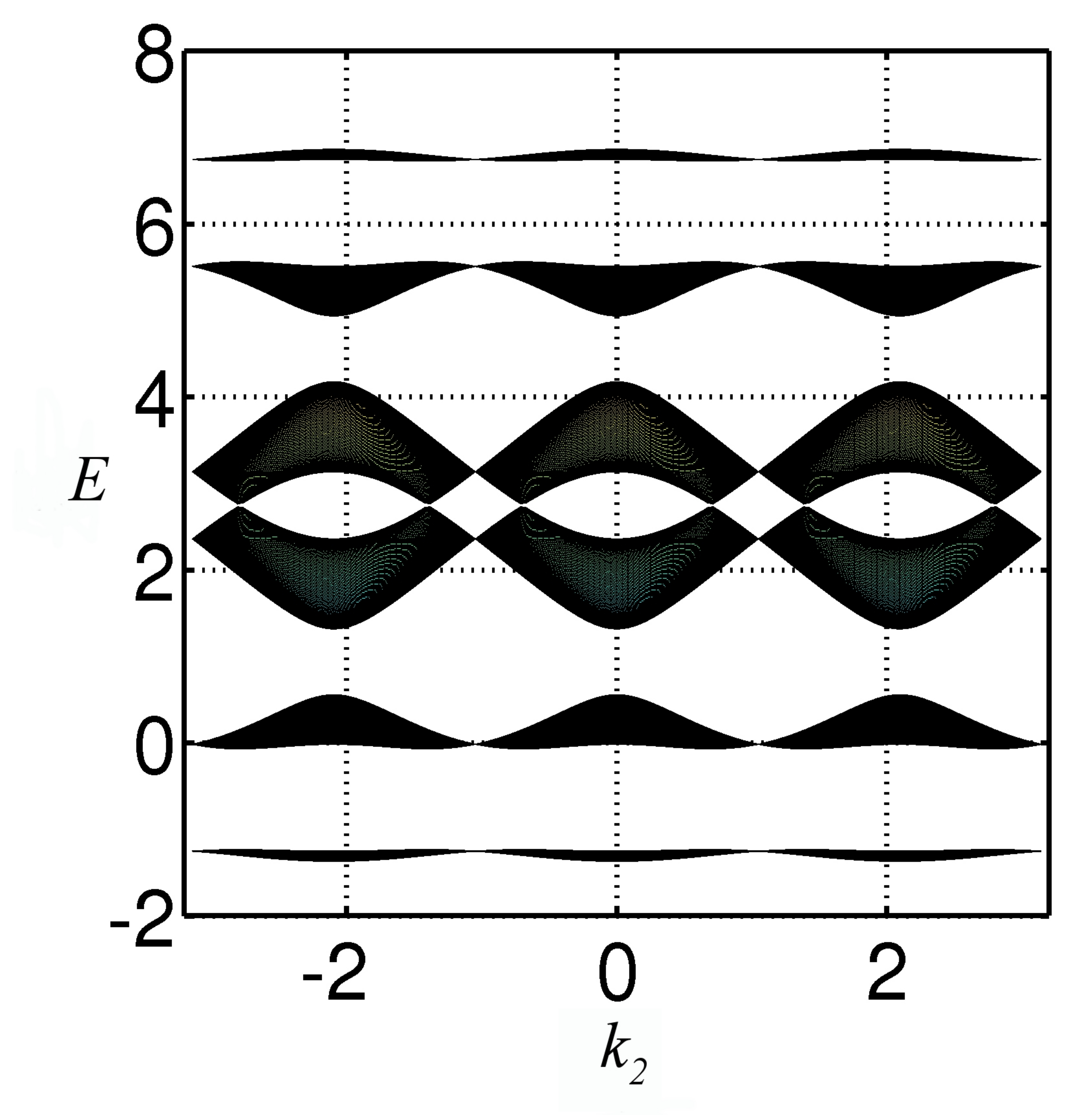}} \hfil
\subfloat[]{\label{q3_n}\includegraphics[scale=0.072]{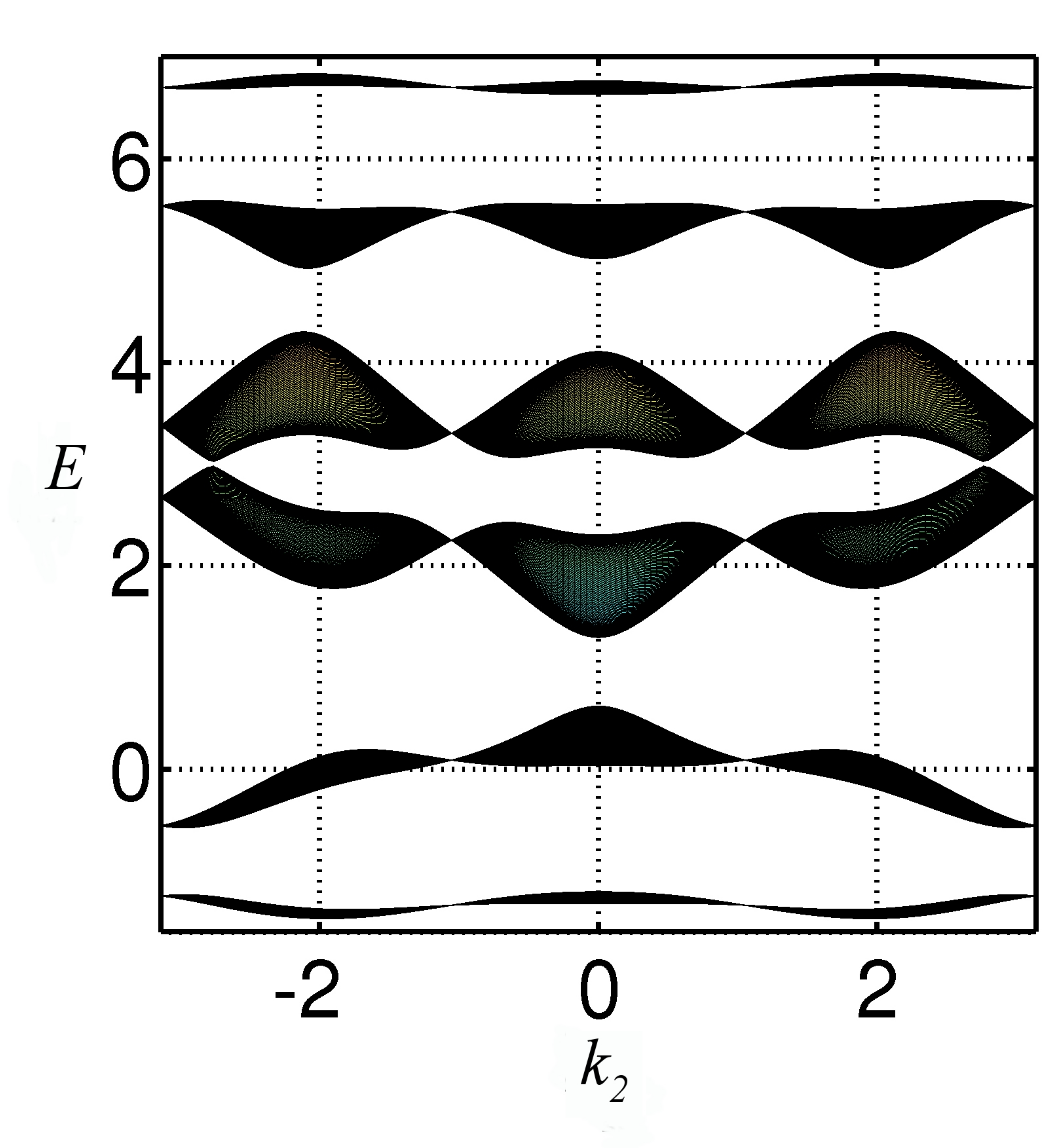}} \hfil
% \subfloat[]{\label{q5}\includegraphics[height=4.5cm, keepaspectratio]{q5_band2_energy1}}
\caption{(Color online) Energy band diagram w.r.t $k_2$   for (a) $q=3$ in symmetric phase for $m=2$ at $V=2.744$, (b) $q=3$ in
  nematic phase for $m=2$ at $V=2.744$. The energy band diagrams are plotted by diagonalizing the
  mean field Hamiltonian using the order parameters obtained by solving the self consistency equations for $m=2$ and $q=3$.}
\end{center}
\end{figure}
Fig.~\ref{q3_s} and Fig.~\ref{q3_n} is the energy band diagram for
$q=3$ at $m=2$ filling in the symmetric and nematic phase respectively
at the transition point $V=2.744$.  We see that in Fig.~\ref{q3_s},
$E_{k_1,k_2}=E_{k_1,k_2+2\pi/3}$ which is the result of the system
being invariant under the translational symmetry unlike in the nematic
phase where translational symmetry is broken which is reflected in the
energy band diagram, Fig.~\ref{q3_n}, where $E_{k_1,k_2}\neq
E_{k_1,k_2+2\pi/3}$.  Hence, in the later case, the length scale, the
periodicity of the mean field Hamiltonian, is now same as that of the
periodicity of the magnetic unit cell. So the ground state and the
periodicity depends on the magnetic field. As seen in Fig.~\ref{q3_s}
and Fig.~\ref{q3_n}, there is always a band gap between the filled
bands and the first excited band (i.e. band gap between between second
and third bands).  The bandwidth of the bands increases in the
translation symmetry broken phase and the band gap decreases but is
never zero in the interaction strength considered. Therefore, the mean
field approximation is a good approximation here.

\end{document}